\documentclass[journal,12pt,onecolumn,draftclsnofoot]{IEEEtran}

\usepackage{graphicx,subfigure}
\usepackage{amsfonts,amsmath,amssymb, mathrsfs, amsthm, cite}
\usepackage{multirow}
\usepackage{epic,eepic,eepicemu}
\usepackage{epsf}
\usepackage{epsfig}
\usepackage{graphics}
\usepackage{array} 
\usepackage{mathrsfs}
\usepackage{psfrag}
\usepackage{tikz}
\usepackage{url}
\usepackage{enumerate}
\usepackage{booktabs}
\usetikzlibrary{arrows}
\usepackage{verbatim}
\usepackage{bm}

\usetikzlibrary{arrows}
\usepackage{verbatim}

\newtheorem{example}{Example}

\newtheorem{proposition}{Proposition}
\newcommand*{\rom}[1]{\expandafter\@slowromancap\romannumeral #1@}
\makeatother

\usepackage[skip=2pt,font=footnotesize]{caption}

\newcommand{\abs}[1]{\lvert#1\rvert}

\newcommand{\be}{\begin{equation}}
\newcommand{\ba}{\begin{align}}
\newcommand{\ee}{\end{equation}}
\newcommand{\ea}{\end{align}}
\newcommand{\ben}{\begin{equation*}}
\newcommand{\een}{\end{equation*}}
\newcommand{\mc}{\mathcal}

\newcommand{\Tc}{\mathcal{T}}
\newcommand{\Vc}{\mathcal{V}}
\newcommand{\Sc}{\mathcal{S}}
\newcommand{\Zc}{\mathbb{Z}}

\newcommand{\syn}{\textsf{syn}}

\newcommand{\summ}{\textsf{sum}}

\DeclareMathSymbol{:}{\mathbin}{operators}{"3A}

\IEEEoverridecommandlockouts

\begin{document}

\title{ Efficient Systematic Encoding of \\ Non-binary VT Codes }
\author{Mahed Abroshan, Ramji Venkataramanan and Albert Guill\'en i F\`abregas\thanks{M. Abroshan and R. Venkataramanan are with the Department of Engineering, University of Cambridge, UK, (ma675@cam.ac.uk, ramji.v@eng.cam.ac.uk).}
\thanks{A. Guill\'en i F\`abregas is with the Department of Information and Communication
Technologies, Universitat Pompeu Fabra, Barcelona 08018, Spain,
also with the Instituci\'o Catalana de Recerca i Estudis Avan\c{c}ats (ICREA),
Barcelona 08010, Spain, and also with the Department of Engineering, University
of Cambridge, Cambridge CB2 1PZ, U.K. (e-mail: guillen@ieee.org).}
\thanks{This work was supported in part by the European Research Council under Grants 259663 and 725411, in part by the Spanish Ministry of Economy and Competitiveness under Grant Grant TEC2016-78434-C3-1-R.}
}

\maketitle
\begin{abstract}
Varshamov-Tenengolts (VT) codes are a class of codes which can correct a single deletion or insertion with a  linear-time decoder. This paper addresses the problem of efficient encoding of non-binary VT codes, defined over an alphabet of size $q >2$.  We propose a simple linear-time encoding method to systematically map binary message sequences onto VT codewords. The method  provides a new lower bound on the size of $q$-ary VT codes of length $n$.
\end{abstract}  

\section{Introduction}

 Designing codes  for correcting deletions or insertions is well known to be a challenging problem; see, e.g.,\cite{Sloane00,DaveyMackay01,ratzer05marker,abdelPFC12,le2016new,brakensiekVZ16,GuruswamiW17,hannaEl17guess}.   
For the special case of correcting one insertion or deletion, there exists an elegant class of codes called Varshamov-Tenengolts (VT) codes. Binary VT codes were first introduced by Varshamov and Tenengolts in \cite{VT65} for  channels with asymmetric errors. Later, Levenshtein \cite{Lev65} showed that they can be used for correcting a single deletion or insertion with a simple decoding algorithm whose complexity is linear in the code length \cite{Sloane00}. Tenengolts subsequently introduced a non-binary version of VT codes, defined over a $q$-ary alphabet for any $q >2$  \cite{Tenengolts84}. The $q$-ary VT codes retain  many of the attractive properties of the  binary codes. In particular, they  can correct deletion or insertion of a single symbol from a  $q$-ary VT codeword with a linear-time decoder.

Given the simplicity of VT decoding, a natural question is: can one construct a linear-time encoder to efficiently map binary  message sequences onto VT codewords? 
For binary VT codes, such an encoder was proposed by Abdel-Ghaffar and Ferriera   \cite{AbdelFer98}. (A similar encoder was also described in  \cite{SKF99}.)  However, the issue of efficient encoding for non-binary VT codes has not been addressed previously, to the best of our knowledge.  In this paper, we propose an efficient systematic encoder for non-binary VT codes. The encoder has complexity that is linear in the code length, and  is systematic in the sense that the message bits are assigned to pre-specified positions in the codeword.  The encoder also yields a new lower bound on the size of  $q$-ary VT codes, for $q >2$.  

VT codes are  a key ingredient of recent code constructions for channels with segmented deletions and insertions \cite{liuMitz10,abroshan2017coding}, and the proposed VT encoder can be applied to these constructions. VT codes have also recently been used in  algorithms for synchronization from deletions and insertions, e.g., \cite{RVsync15,YazdiDol14,sala16sync,abroshan2017multilayer}. 

In \cite[Sec. 5]{Tenengolts84},  Tenengolts  introduced a  systematic non-binary code that can  correct a single deletion or insertion. However, this code is not strictly a VT code as its codewords do not necessarily share the same VT parameters. (Formal definitions of VT codes and their parameters are given in the next two sections.) In this paper, we propose an encoder for VT codes defined in the standard way, noting that using standard VT codes is a key requirement in some of the code constructions mentioned above.  

\emph{Notation}: Sequences are denoted by capital letters, and scalars by lower-case letters. Throughout, we use $n$ for the code length and $k$ for the number of message bits mapped to the codeword. The set $\mathbb{Z}_q=\{0,1,\cdots,q-1\}$ is the finite integer ring of size $q$. We consider the natural order for the elements of $\mathbb{Z}_q$, i.e., $0 <1 \ldots <(q-1)$.  The term dyadic index will be used to refer to an index that is a power of $2$.

The rest of the paper is organized as follows. In the next section, we formally define binary VT codes and briefly review the systematic encoder from \cite{AbdelFer98}.  In Section \ref{nonbinary-sec}, we define the $q$-ary VT codes, and describe the systematic encoding method and the resulting lower bound on the size of the codes.


\section{Binary VT codes} \label{sec:VT}
The VT syndrome of a binary sequence $S = s_1s_2\cdots s_n \in \mathbb{Z}_2^n$  is defined as 
\begin{equation}
\syn(S)\triangleq \sum_{i=1}^n i \,s_i \  \  \text{ mod }  (n+1).
\label{eq:VTformula}
\end{equation}
For positive integers $n$ and $0\leq a\leq n$, the VT code of length $n$ and syndrome $a$, is defined as 
\be
\Vc\Tc_a(n) = \big\{S \in \mathbb{Z}_2^n: \syn(S)=a\big\}, 
\ee
i.e., the set of binary sequences $S$ of length $n$ that satisfy $\syn(S)=a$.  For example, the VT code of length $3$ and syndrome $2$ is
\begin{align}
\begin{split}
\Vc\Tc_2(3) &=  \Bigl\{s_1s_2s_3\ \in\mathbb{Z}_2^3 :\sum_{j=1}^3 j\,s_j = 2 \text{ mod 4} \Bigr\}\\
&=\{010, 111 \}.
\end{split}
\end{align}

Each of the sets $\Vc\Tc_a(n)$,  $0\leq a\leq n$,  is  a code that can correct  a single deletion or insertion with a decoder whose complexity is linear in  $n$. The details of the decoding algorithm can be found in  \cite{Sloane00}.

The $(n+1)$ sets $\Vc\Tc_a(n)$,  $0\leq a\leq n$, partition the set of all binary sequences of length $n$, i.e., each sequence $S\in\Zc_2^n$ belongs to exactly one of the sets. Therefore, the smallest of the  codes $\Vc\Tc_a(n)$ will have at most $\frac{2^n}{n+1}$ sequences. Hence
\be
 \min_{0 \leq a \leq n} \ \frac{1}{n} \log_2\vert \Vc\Tc_{a}(n)\vert \leq 1 -  \frac{1}{n}\log_2(n+1).
 \label{eq:VTRmin}
\ee
An exact formula for the size  $\abs{ \Vc\Tc_a(n)}$ was given by Ginzburg \cite{ginzburg67}. The formula  \cite[Theorem 2.2]{Sloane00} does not give an analytical expression for general $n$, but it shows that if $(n+1)$ is a power of $2$,  then $\vert \Vc\Tc_a(n) \vert = 2^n/(n+1)$, for $0 \leq a \leq n$. 
Moreover, for general $n$, the formula can be used to deduce that the sizes of the  codes $\Vc\Tc_a(n)$ are all approximately $2^n/(n+1)$. In particular, 
\be
\frac{2^n}{(n+1)}  - 2^{(n+1)/3}  \leq  \vert \Vc\Tc_a(n) \vert  \leq \frac{2^n}{(n+1)}  + 2^{(n+1)/3}, \quad \ \text{ for } a \in \{0, \ldots, n \}.
\label{eq:sloane_bnds}
\ee

Abdel-Ghaffar and Ferriera   \cite{AbdelFer98} proposed a systematic encoder  to map $k$-bit message sequences onto codewords in $\vert \Vc\Tc_a(n) \vert$, where $k=n - \lceil\log_2 (n+1)\rceil$.  
We briefly review the encoding procedure as it is an ingredient of our systematic $q$-ary VT encoder.  Consider a $k$-bit message $M=m_1\cdots m_{k}$ to be encoded into a codeword $C = c_1\cdots c_n\in\Vc\Tc_a(n)$, for some $a\in \{0,1,\cdots,n\}$. The number of ``parity" bits  is denoted by $t = n-k =\lceil\log_2 (n+1)\rceil$. The  idea is to use the code bits in dyadic positions, i.e., $c_{2^i}$, for $0\leq i\leq (t-1)$,  to ensure that $\syn(C) =a$.  The encoding steps are:

\begin{enumerate}
\item Denote the first $k$ non-dyadic indices by $\{ {j_1},\cdots, {j_{k}}\}$, where the indices are in ascending order, i.e., $j_1=3, j_2=5, \ldots$ 
We set $c_{j_i}$ equal to the  message bit $m_i$, for $1\leq i \leq k$.

 \item First set the bits in all the dyadic positions to be zero and denote the resulting sequence by $C'=c'_1\cdots c'_n$ (so that we have $c'_{2^i}=0$ for $0\leq i\leq t-1$ and $c'_{j_l}=m_l$ for $1\leq l \leq k=(n-t)$). Define the deficiency $d$ as the difference between the desired syndrome $a$ and the syndrome of $C'$. That is, 
\be
d=a-\syn(C') \hspace{3mm} \text{mod } (n+1).
\ee

\item Let the binary representation of $d$ be $d_{t-1} \ldots d_1 d_0$, i.e., $d=\sum_{i=0}^{t-1}2^i d_i$. Set $c_{2^i}=d_i$, for $0 \leq i \leq (t-1)$, to obtain $C$.
\end{enumerate}

The rate of this systematic encoder is $R = 1 - \frac{1}{n} \lceil \log_2(n+1) \rceil$, regardless of the syndrome  $ a \in  \{0, \ldots, n \}$. Comparing with \eqref{eq:VTRmin}, we observe  that the rate loss for the smallest VT code of length $n$ is less than $\frac{1}{n}$. On the other hand, if $(n+1)$ is  not a power of two, the rate loss for the larger VT codes may be higher due to  codewords that are unused  by the  encoder. However this rate loss is unavoidable with any systematic encoder \cite{AbdelFer98}.

We remark that the dyadic positions are not the only set of positions that can be used for syndrome bits. For instance, the following set of indices also produce all syndromes:
$$\{c_{i_0},\cdots,c_{i_{t-1}}\}\quad\text{where}\quad i_j=-2^j \quad (\text{mod }n+1)\quad \text{for } 0\leq j\leq t-1.$$
This can be helpful in some applications (see e.g. \cite{abroshan2017coding}) where some code bits are already reserved for  prefixes or suffixes,  and thus cannot be used as syndrome bits. In general, a set of positions $\{p_1,p_2,\cdots,p_r\}$ can be used for syndrome bits if for each syndrome $a\in \{0,\cdots,n\}$, there exists a subset $\mathcal P\subseteq \{p_1,p_2,\cdots,p_r\}$ such that 
\be \sum_{j\in \mathcal P} p_{j}=a \hspace{3 mm} (\text{mod } n+1). \ee
In other words, for each $a\in \{0,\cdots,n\}$,  there should exist binary coefficients $b_1,\cdots,b_r$ such that  
\be \sum_{j=1}^r b_jp_j=a \quad (\text{mod } n+1).\ee

\section{Efficient Encoding for Non-binary VT codes}\label{nonbinary-sec}

For any code length $n$, the VT codes over $\Zc_q$, for $q>2$ are defined  as follows \cite{Tenengolts84}.  For each $q$-ary sequence $S=s_0s_1\cdots s_{n-1}\in \Zc_q^n$, define a corresponding length $(n-1)$ auxiliary binary sequence $A_S=\alpha_1\alpha_2\dotsc\alpha_{n-1}$  as follows\footnote{For non-binary sequences, we start the indexing from $0$ as this makes it convenient to describe the encoding procedure in Section \ref{nonbinary-sec}.}. For $1 \leq i \leq n-1$,
\be\label{aux} 
\alpha_i =
  \begin{cases}
    1  & \quad \text{if } s_i\geq s_{i-1}\\
    0  & \quad \text{if } s_i< s_{i-1}. \\
  \end{cases}
\ee
We also define the modular sum of $S$ as
\be
\summ(S)=\sum_{i=0}^{n-1} s_i \quad (\text{mod } q).
\ee
For $0\leq a\leq n-1$ and $b\in \mathbb{Z}_q$, the $q$-ary VT code with length $n$ and  parameters $(a,b)$ is defined as 
\be
\Vc\Tc_{a,b}(n) = \big\{S \in \mathbb{Z}_q^n: \syn(A_S)=a,\, \summ(S)=b\big\}.
\ee
Each of the sets $\Vc\Tc_{a,b}(n)$  is  a code that can correct  deletion or insertion of a single symbol with a decoder whose complexity is linear in the code length $n$. The details of the decoding algorithm can be found in \cite[Sec. II]{Tenengolts84}.

 Similarly to the binary case, the codes $\Vc\Tc_{a,b}(n)$,  for $0\leq a\leq n-1$ and $b\in \mathbb{Z}_q$, partition the space $\mathbb{Z}_q^n$ of all $q$-ary sequences of length $n$. For a given $n$, there are $nq$ of these codes, and hence the smallest of them will have no more than $\frac{q^n}{nq}$  sequences.  Let $R_{\text{min}}$ be the rate of the smallest of these  codes, i.e., 
\begin{align}
R_{\text{min}}&\triangleq \min_{a,b} \, \frac{\log_2 \vert \Vc\Tc_{a,b}(n)\vert}{n},
\end{align}
where the minimum is over $0\leq a\leq n-1$ and $b\in \mathbb{Z}_q$.
We then have the bound 
\begin{align}
R_{\text{min}} & \leq \log_2 q-\frac{1}{n}\log_2 n-\frac{1}{n}\log_2 q  \ \  \text{bits/symbol}. \label{rate-vt-nb}
\end{align}
The encoding procedure described below yields a lower bound on the size of $\vert \Vc\Tc_{a,b}(n)\vert$ (see Proposition \ref{prop:qLB}), which shows that for $q \geq 4$,
\be
R_{\text{min}} \geq  \log_2 q-\frac{1}{n} \lceil \log_2 n \rceil (3 \log_2 q - 2 \log_2 (q-1) )-\frac{1}{n}( 5 \log_2 (q-1) - 3 \log_2 q )
\ \  \text{bits/symbol}.
\ee
Kulkarni and Kiyavash \cite{KK13} have shown that  the size of any single deletion correcting $q$-ary code of length $n$ is bounded by $\frac{q^n -q}{(q-1)(n-1)}$. This yields a rate upper bound $R_{\text{max}}$ for any single deletion correcting code, where
\be
R_{\text{max}} \leq  \log_2 q-\frac{ \log_2 (n-1)}{n} - \frac{\log_2 (q-1)}{n}.
\label{eq:Rmax}
\ee

We now describe the encoder to map a sequence of message bits to a  codeword of the $q$-ary VT code $\Vc\Tc_{a,b}(n)$.
For simplicity, we first assume that $q$ is a power of two, and address the case of  general $q$ at the end of this section. 
We will map a $k$-bit message $M=m_1m_2\cdots m_k$ to a codeword in $\Vc\Tc_{a,b}(n)$, where
\begin{align}
k&=(n-3t+3)\log_2 q+(t-3)(2\log_2 q-1)+(\log_2 q-1) \label{eq:nb_three_terms}\\
 &=n\log_2 q-t(\log_2 q+1)-2(\log_2 q-1),  \label{k-nb}
\end{align} 
with $t=\lceil\log_2 n\rceil$.  Therefore,  the rate of our encoding scheme is
\begin{align}
R=\log_2 q -\frac{\lceil\log_2 n\rceil (\log_2 q+1)}{n}-\frac{2\log_2 q-2}{n} \ \ \text{bits/symbol}.
\end{align}

Our encoding method gives a  lower bound on the size of any non-binary VT code of length $n$.  An immediate lower bound on  the size is $2^k$, with $k$  given by \eqref{k-nb}.  The proposition below gives a slightly better bound, which  is obtained  by  modifying the encoding method to map $q$-ary message sequences to  $q$-ary VT codewords, rather than  a binary message sequence to a $q$-ary VT codeword.
\begin{proposition}
For $n\geq 6$, $q\geq 4$, and any  $0\leq a\leq n$, and $b\in \Zc_q$, we have 
\begin{align*}  \vert \Vc\Tc_{a,b}(n)\vert & \geq (q-1)^{2t-5}q^{n-3t+3}, \label{lower-nb} \\
& = q^{n\left(1 - \frac{t}{n}[ 3 \log_2 q - 2 \log_2 (q-1)] - \frac{1}{n} [5 \log_2(q-1) - 3 \log_2 q ] \right)} \end{align*}
where $t=\lceil\log_2 n\rceil$.
\label{prop:qLB}
\end{proposition}
  The proof of the proposition is given at the end of this section, after describing the encoding procedure.  We emphasize that  we use $t=\lceil\log_2 n\rceil$ throughout this section (as opposed to  $\lceil\log_2 (n +1) \rceil$ used for binary VT encoding) because the binary auxiliary sequence has length $(n-1)$.

\subsection{Encoding procedure}
The high level idea for mapping a $k$-bit message to a codeword $C\in \Vc\Tc_{a,b}(n)$ is the following. Similarly to the binary case, we reserve the $t$ dyadic positions in the  binary auxiliary sequence $A_C$ to ensure that $\syn(A_C)=a$. Recall from \eqref{aux} that each bit of $A_C$ is determined by comparing two adjacent symbols of the $q$-ary sequence $C$. Therefore, to ensure that $\syn(A_C)=a$, in addition to reserving  the symbols in the dyadic positions of $C$, we also place some restrictions on the symbols adjacent to the dyadic positions. Finally, we use the first three symbols of $C$ to ensure that $\summ(C)=b$. We explain the method in six steps with the help of the following running example.
\begin{example}\label{exm-2}
Let $q=8, n=16$, and suppose that we wish to encode a binary message $M$ to a codeword $C$ in $\Vc\Tc_{0,1}(16)$. We have $t=\lceil\log_2 n \rceil=4$ and $\log_2 q=3$. Therefore, from \eqref{eq:nb_three_terms} the length of $M$  is $k = 3(n-3t+3)+5(t-3)+2=28$ bits. Let 
\be M=110\ 001\ 000 \ 111 \ 010 \ 101\ 000 \ 11100 \ 11, \ee
where the spacing indicates the bits corresponding to the three terms in \eqref{eq:nb_three_terms}.
\end{example}
\textbf{Step 1.} 
Let $\mathcal S$ be the set of pairs of symbols adjacent to a dyadic symbol, i.e., 
\be \Sc=\{(c_{2^j-1},c_{2^j+1}), \ \text{ for } 2\leq j\leq (t-1)\}. \ee 
There are $\vert \Sc\vert=(t-2)$ pairs of symbols in $\Sc$.
Excluding $c_0$, the number of symbols in $C$ that are  neither in dyadic positions nor in $\Sc$  is \be (n-1)-2\vert\mathcal S\vert-t=(n-3t+3). \ee  Assign the first $(n-3t+3)\log_2 q$ bits of the message $M$ to these symbols, by converting each set of $\log_2 q$ bits to a $q$-ary symbol. This corresponds to the first term in \eqref{eq:nb_three_terms}.

In Example \ref{exm-2}, $(n-3t+3)\log_2 q =21$, and  the representation of first $21$ bits of $M$ in $\Zc_8$ is  $6\ 1\ 0\ 7\ 2\ 5\ 0$.  Therefore the sequence $C$ is
\be C= c_0\ c_1\ c_2\ c_3\ c_4\ c_5\ 6\ c_7\ c_8\ c_9\ 1\ 0\ 7\ 2\ 5\ 0. \label{eq:step1_example} \ee

\textbf{Step 2.} 
In this step, we assign the remaining bits of the message to the symbols in $\Sc$. For a given dyadic position $c_{2^j}$, $j=2,3,\cdots,(t-1)$, we constrain the pair of adjacent symbols $(c_{2^j-1},c_{2^j+1})$ to belong to the following set
\be \label{eq:Tcdef} \Tc=\{(r,l)\in \mathbb{Z}_q\times \mathbb{Z}_q : r\neq 0, \ l \neq (r-1) \}. \ee
Via \eqref{eq:Tcdef}, we enforce  $c_{2^j-1} \neq 0$ because if $c_{2^j-1} $ were $0$, then we necessarily have $c_{2^j}\geq c_{2^j-1}$ which constrains the value of $\alpha_{2^j}$ to $1$. Recall from \eqref{aux} that $\alpha_1\ldots \alpha_{n-1}$ is the auxiliary sequence. However, $\alpha_{2^j}$ needs to be unconstrained in order to guarantee that any desired syndrome can be generated. Furthermore, we will  see in Step 5 that if $c_{2^j+1}=c_{2^j-1}-1$, then we may be unable to find a suitable symbol $c_{2^j}$ compliant with the restrictions induced by the auxiliary sequence. We therefore enforce the constraint $c_{2^j+1} \neq c_{2^j-1}-1$ using  \eqref{eq:Tcdef}.   It is easy to see that $\vert\mathcal T\vert=(q-1)^2$.
 
  Excluding the pair $(c_3,c_5)$, there are $(t-3)$ pairs in $\Sc$. If we were encoding $q$-ary message symbols, each of these $(t-3)$ pairs could take any pair of symbols in $\mc{T}$.  Since we are encoding message bits, we  use a look up table to map $\lfloor\log_2 \abs{\mc{T}} \rfloor=2\log_2 q-1$ bits to each of the pairs in $\Sc$ excluding $(c_3,c_5)$. We thus map $(t-3)(2\log_2 q-1)$ bits to the pairs in $\Sc$ excluding $(c_3,c_5)$. This corresponds to the second term in \eqref{eq:nb_three_terms}. 
    
  Next,  set $c_3=q-1$. This choice is important as it will facilitate step 6. Since $(c_3,c_5)\in\Tc$, when $c_3=q-1$, then $c_5$ has to be such that $c_5\neq q-2$. Hence, there are $q-1$ possible values for $c_5$. As we are encoding a binary message, we map $\lfloor\log_2 (q-1)\rfloor=\log_2 q-1$ bits to $c_5$ using a look-up table. This corresponds to the third term in \eqref{eq:nb_three_terms}.  We note that the two  look-up tables used in this step have sizes at most $(q-1)^2$ and $q$, respectively. Thus, in steps one and two in total we have mapped the claimed $k$ message bits to  the symbols of $C$.  

In Example \ref{exm-2}, as seen from \eqref{eq:step1_example}, $(c_7,c_9)$ is the only pair in $\Sc$ other than $(c_3,c_5)$.  We  can assign $2\log_2 q-1=5$ bits to $(c_7,c_9)$. After the first 18 message bits  mapped in Step 1, the next five bits in $M$ are $11100$. Suppose that in our look-up table these bits correspond to the pair $(3,5)$. We then have $(c_7,c_9)=(3,5)$. Also, we fix $c_3=q-1=7$, and the last two message bits  determine $c_5$. The last two message bits are $11$. Suppose that $3$ is the corresponding symbol in the look-up table. We therefore set $c_5=3$. Therefore, we have
\be C= c_0\ c_1\ c_2\ 7\ c_4\ 3\ 6\ 3\ c_8\ 5\ 1\ 0\ 7\ 2\ 5\ 0. \label{eq:ex_step2} \ee
Up to this point, we have mapped our $k$ message bits to a partially filled $q$-ary sequence. In the following steps we ensure that the resulting sequence lies in the correct VT code by carefully choosing remaining $(t+1)$ symbols to obtain the  auxiliary sequence syndrome $a$ and the modular sum $b$.

\textbf{Step 3.}
In this step, we specify the bits in the non-dyadic locations of the auxiliary sequence $A_C$. Notice that according to \eqref{aux}, in order to define $\alpha_{2^j+1}$, the value of $c_{2^j}$ should be known. This is not the case here as the dyadic positions in $C$ have been reserved to generate the required syndrome. To circumvent this issue, we determine $\alpha_{2^j+1}$ (for $1<j<t$) by comparing $c_{2^j+1}$ with $c_{2^j-1}$ as follows:
\be 
  \alpha_{2^j+1} = \begin{cases}   1 & \quad \text{if } c_{2^j+1}\geq c_{2^j-1}, \\
  0 & \quad \text{if } c_{2^j+1}< c_{2^j-1}.\\
  \end{cases}
  \label{aux_1}
\ee 
As we shall  show in step 5, we will be able to make these choices for the auxiliary sequence compatible with the definition of a valid auxiliary sequence in \eqref{aux}.

Next, since we have chosen $c_3=q-1$, from the rule in \eqref{aux} we have $\alpha_3=1$, regardless of what $c_2$ is. The other non-dyadic positions of the auxiliary sequence $A_C$ can be filled in using  \eqref{aux}, i.e., $\alpha_i =1$ if $c_i  \geq c_{i-1}$, and $0$ otherwise. 

For our example with $C$ shown in  \eqref{eq:ex_step2}, at the end of this step  we have
\be A_C=\alpha_1\ \alpha_2\ 1\ \alpha_4\ 0\ 1\ 0\ \alpha_8\ 1\ 0\ 0\ 1\ 0\ 1\ 0. \label{exm-aux}\ee

\textbf{Step 4.}
In this step, we use the binary encoding method described in Section \ref{sec:VT} to find the bits in the dyadic positions $\alpha_{2^0},\cdots,\alpha_{2^{t-1}}$ such that $\syn(A_C)=a$. With this, the auxiliary sequence $A_C$ is fully determined.

In the example, we need to find $\alpha_1,\alpha_2,\alpha_4$ and $\alpha_8$ such that $\syn(A_C)=0$. First, we set the syndrome bits $\alpha_1=\alpha_2=\alpha_4=\alpha_8=0$, and denote the resulting sequence by 
\be 
A'_C=0\ 0\ 1\ 0\ 0\ 1\ 0\ 0\ 1\ 0\ 0\ 1\ 0\ 1\ 0.
\ee 
Now, $\syn(A'_C)=12$, and the deficiency  $d=0-12 \,(\text{mod } 16) = 4$. The binary representation  of $d$ is $d_3d_2d_1d_0=0100$. Hence, $\alpha_1=\alpha_2=\alpha_8=0$ and $\alpha_4=1$ will produce the desired syndrome.  Summarizing, we have the following auxiliary sequence
\be A_C=0\ 0\ 1\ 1\ 0\ 1\ 0\ 0\ 1\ 0\ 0\ 1\ 0\ 1\ 0. \label{eq:ex_Ac_complete} \ee

\textbf{Step 5.} In this step, we specify the symbols of $C$ in the dyadic positions (except $c_1$ and $c_2$). This will be done by ensuring that $A_C$ is a valid auxiliary sequence consistent with the definition in \eqref{aux}. In particular, the choice of $c_{2^j}$ for $j=2,\cdots,t-1$, should be consistent with $\alpha_{2^j+1}$ and $\alpha_{2^j}$. We ensure this by choosing $c_{2^j}$ (for $1<j<t$) as follows:
\be \label{rule}
c_{2^j} = \begin{cases}   
    c_{2^j-1}-1 & \quad \text{if } \alpha_{2^j}=0, \\
    c_{2^j-1} & \quad \text{if } \alpha_{2^j}=1.\\
\end{cases}
\ee
From the definition in \eqref{aux}, this choice is consistent with $\alpha_{2^j}$. Now we show that it is also consistent with $\alpha_{2^j+1}$. If $\alpha_{2^j+1}=1$, then according to \eqref{aux_1}, $c_{2^j+1}\geq c_{2^j-1}$; then the choice of  $c_{2^j}$ in \eqref{rule} always guarantees that $c_{2^j-1}\geq c_{2^j}$, and thus $c_{2^j+1}\geq c_{2^j}$.  Next suppose that $\alpha_{2^j+1}=0$. Then according to \eqref{aux_1}, $c_{2^j+1}< c_{2^j-1}$.  We need to verify that $c_{2^j+1} < c_{2^j}$ Now, if $\alpha_{2^j}=1$, then  $c_{2^j}= c_{2^j-1}$ and $c_{2^j+1}< c_{2^j}$. Also, if $\alpha_{2^j}=0$, from \eqref{rule} we have $c_{2^j}= c_{2^j-1}-1$. Since symbols adjacent to dyadic positions $(c_{2^j-1},c_{2^j+1})$ are chosen from $\Tc$ (see step 2), then $c_{2^j+1}\neq c_{2^j-1}-1$. Thus, we have that $c_{2^j+1}<c_{2^j-1}-1=c_{2^j}$. Therefore, in either case the choice is consistent with \eqref{aux}.

For the example, using \eqref{rule}  and \eqref{eq:ex_Ac_complete} we obtain
\be C=c_0\ c_1\ c_2\ 7\ 7\ 3\ 6\ 3\ 2\ 5\ 1\ 0\ 7\ 2\ 5\ 0. \label{eq:c3_complete} \ee

\textbf{Step 6.}
Finally, we need to find $c_0,c_1$ and $c_2$ that are compatible with $\alpha_1,\alpha_2,\alpha_3$ (the first three bits of the auxiliary sequence), and such that $\summ(C)=b$. Let
\be w\triangleq b-\sum_{i=3}^n c_i \quad (\text{mod } q).\label{w-def}\ee 
Hence we need $c_0+c_1+c_2=w  \text{ (mod } q\text{)}$. We will show that when $q\geq 4$, we can find three  distinct integers $(x,y,z)$ such that $0\leq x<y<z <q$ and $x+y+z=w \text{ (mod } q\text{)}$.  
We will assign these numbers to $c_0,c_1$ and $c_2$. 
Also recall that we set $c_3=q-1$; hence we always have $x,y,z\leq c_3$,  which is consistent with $\alpha_3=1$.

The triplet with smallest numbers that we can choose is $x=0,y=1,z=2$. For this choice, $w=0+1+2=3\text{ (mod } q\text{)}$. By increasing $z$ from $2$ to $q-1$ with $x=0$ and $y=1$, we can produce any value of $w$ from $3$ to $q-1$ as well as $w=0$.  Finally, the only remaining values are $w=1,2$. To obtain these values, we choose $x,y,z$ as follows.
\begin{enumerate}
\item $\underline{w=1}$: Choose $x=0, y=2, z=q-1$.
\item $\underline{w=2}$: Choose $x=1, y=2, z=q-1$.
\end{enumerate}
 Hence, we have shown that for $q \geq 4$,  any $w\in\mathbb{Z}_q$ can be expressed as the (mod $q$) sum of three distinct elements of $\Zc_q$. Assigning these elements to $c_0, c_1, c_2$ in the order required by the auxiliary sequence completes the encoding procedure. We now have $\summ(C)=b$ and $\syn(A_C)=a$, and thus $C\in \Vc\Tc_{a,b}(n)$ as required.

In our example, from \eqref{eq:c3_complete} we have 
\be\sum_{i=3}^{15} c_i=48=0 \quad (\text{mod }8),\ee
and $b=1$.  Therefore we need $c_0+c_1+c_2=1 \text{ (mod } 8\text{)}$. 
We have $\alpha_1=\alpha_2=0$ so $c_0>c_1>c_2$ is the correct order. We therefore assign $c_0=7$, $c_1=2$, and $c_2=0$ to obtain the codeword. 
\be C= 7\ 2\ 0\ 7\ 7\ 3\ 6\ 3\ 2\ 5\ 1\ 0\ 7\ 2\ 5\ 0\in \Vc\Tc_{0,1}(16). \ee
It can be verified that   $\summ(C)=1$, and the auxiliary sequence syndrome $\syn(A_C)=0$.

\subsection{The case where $q$ is not a power of two} \label{subsec:gen_q}

When $\log_2 q$ is not an integer, the main difference is that we map longer sequences of bits to sequences of $q$-ary symbols. Recall that in step $1$,  we determine $(n-3t+3)$ symbols of the $q$-ary codeword.  One can map $\lfloor(n-3t+3)\log_2 q\rfloor$ bits to these $(n-3t+3)$ symbols using standard  methods to convert an integer expressed  in base $2$ into base $q$. In the second step, as described earlier we can map $\lfloor\log_2 (q-1)^2\rfloor$ bits to $(t-3)$ pairs in $\Sc$ (excluding $(c_3,c_5)$). Moreover $\lfloor\log_2 (q-1)\rfloor$ bits can be mapped to $c_5$. Therefore, in total we can map $k$ bits to a $q$-ary VT codeword of length $n$, where 
\begin{align}
k&=\lfloor(n-3t+3)\log_2 q\rfloor+(t-3)\lfloor\log_2 (q-1)^2\rfloor+\lfloor\log_2 (q-1)\rfloor\\
&\geq n\log_2 q-t(\log_2 q+2)-(2\log_2 q-4).
\end{align} 
For $q\geq 4$, the remaining steps are identical to the case where $q$ is a power of two.
The case of $q=3$ is slightly different and it is discussed in Appendix \ref{app:q3proof}.

\subsection{Proof of Proposition 1}
The result can be directly derived from steps one and two of our encoding method by mapping  sequences of $q$-ary message symbols (rather than sequences of message bits) to distinct codewords in $\vert \Vc\Tc_{a,b}(n)\vert$. In step 1, we can assign $(n-3t+3)$ arbitrary symbols to positions that are neither dyadic nor in $\Sc$. There are $q^{n-3t+3}$ ways to choose these symbols. Then in step two, we can choose $(q-1)^2$ pairs for each of the $(t-3)$ specified pairs of positions; furthermore, there are $(q-1)$ choices for $c_5$.  According to steps 3 to 6, we can always choose the  remaining symbols such that resulting codeword lies in $\Vc\Tc_{a,b}(n)$. Therefore,  we can  map $q^{n-3t+3}(q-1)^{2t-5}$ different sequences of message symbols to distinct codewords in $\Vc\Tc_{a,b}(n)$. This yields the lower bound on $\vert \Vc\Tc_{a,b}(n)\vert$.

\appendix
\subsection{Encoding for $q=3$} \label{app:q3proof}
For $q=3$, we need to slightly modify the proposed algorithm. The first step is as described in Section \ref{subsec:gen_q}. The difference in the second step  is that  we do not embed data in $c_5$ and simply choose $c_5=c_3=2$. Steps three to five remain the same. In the sixth step, we  compute $w$ as in \eqref{w-def}, and choose $c_0,c_1, c_2$ as follows depending on the values of $\alpha_0$ and $\alpha_1$:
\begin{enumerate}
\item $\underline{\alpha_1=\alpha_2=1}$: Choose $c_2=c_1=2$ and $c_0=w-4 \text{ (mod }3)$.
\item $\underline{\alpha_1=1, \alpha_2=0}$: Choose $c_2=1$ and $c_1=2$ and $c_0=w-4 \text{ (mod }3)$.
\item $\underline{\alpha_1=0, \alpha_2=1}$: Choose $c_2=2$. If $w=1$, then $c_1=0, c_0=2$. If $w=0$, then $c_1=0, c_0=1$. If $w=2$, then $c_1=1, c_0=2$.
\end{enumerate}   
The only remaining case is when $\alpha_1=\alpha_2=0$. For this case, we need to change $c_3$ and $c_4$, and also the first three bits of $A_C$.  Since $c_3$ has been set to $2$, the first three bits of $A_C$ in this case are $001$. If we change these three bits to $110$, $\syn(A_C)$ will remain unchanged. We therefore set $\alpha_1=\alpha_2=1$ and $\alpha_3=0$. Now we update  $c_0,c_1,c_2,c_3$ to be compatible with the new auxiliary sequence. Set $c_3=1$, recall that $c_5=2$ so  we still have $c_5\geq c_3$ and hence this change will not affect $\alpha_5$. Update $c_4$ according to \eqref{rule}. Set $c_2=c_1=2$, and $c_0=w-4 \text{ (mod }3)$. Now we have $c_3<c_2$ which is consistent with $\alpha_3=0$. Also $c_2\geq c_1 \geq c_0$ is consistent with $\alpha_1=\alpha_2=1$.

Hence, for $q=3$, we have mapped $k=\lfloor\log_2 3(n-3t+3)\rfloor+2(t-3)$ bits to a $q$-ary codeword $C$. This  induces following rate:
\begin{align}
R&=\frac{\lfloor\log_2 3\ (n-3\lceil\log_2 n\rceil+3)\rfloor}{n}+\frac{2(\lceil\log_2 n\rceil-3)}{n}\\
&\geq \log_2 3 -\frac{2.76\ \lceil\log_2 n\rceil}{n} - \frac{2.25}{n}
\end{align} 
Similarly to Proposition 2, we can show that for $q=3$ there are at least $2^{2(t-3)}3^{n-3t+3}$ codewords in each of the VT codes.

\subsection*{Acknowledgement}

The authors thank Andreas Lenz for pointing them to  the references \cite{AbdelFer98} and \cite{SKF99}.

\bibliographystyle{ieeetr}
\bibliography{ref}
\end{document}